# Evolving Deep Convolutional Neutral Network by Hybrid Sine-Cosine and Extreme Learning Machine for Real-time COVID19 Diagnosis from X-Ray Images

Wu Chao[1], Mohammad Khishe[2*], Mokhtar Mohammadi[3], Sarkhel H. Taher Karim[4], Tarik A. Rashid[5]

[1]Department of Radiology, The First Affiliated Hospital of Chongqing Medical University, China.

[2*]Corresponding Author, Imam Khomeini Marine Science University, Nowshahr, Iran, m_khishe@iust.alumni.ac.ir

[3]Department of Information Technology, Lebanese French University, Erbil, KRG, Iraq.

[4] Computer Department, College of Science, University of Halabja, Halabja, Iraq.

[5]Computer Science and Engineering Department, Science and Engineering Science, University of Kurdistan Hewler, Erbil, KRG, Iraq.

**Abstract**

The COVID19 pandemic globally and significantly has affected the life and health of many communities. The early detection of infected patients is effective in fighting COVID19. Using radiology (X-Ray) images is perhaps the fastest way to diagnose the patients. Thereby, deep Convolutional Neural Networks (CNNs) can be considered as applicable tools to diagnose COVID19 positive cases. Due to the complicated architecture of a deep CNN, its real-time training and testing become a challenging problem. This paper proposes using the Extreme Learning Machine (ELM) instead of the last fully connected layer to address this deficiency. However, the parameters' stochastic tuning of ELM's supervised section causes the final model unreliability. Therefore, to cope with this problem and maintain network reliability, the sine-cosine algorithm was utilized to tune the ELM's parameters. The designed network is then benchmarked on the *COVID-Xray-5k* dataset, and the results are verified by a comparative study with canonical deep CNN, ELM optimized by cuckoo search, ELM optimized by genetic algorithm, and ELM optimized by whale optimization algorithm. The proposed approach outperforms comparative benchmarks with a final accuracy of 98.83% on the *COVID-Xray-5k* dataset, leading to a relative error reduction of 2.33% compared to a canonical deep CNN. Even more critical, the designed network's training time is only 0.9421 milliseconds and the overall detection test time for 3100 images is 2.721 seconds.

## 1. Introduction

In recent decades, the detection and diagnosis of various diseases have been successfully investigated by scientists [1-5]. However, the early diagnosis of coronavirus has become a challenge for scientists due to the limited treatments and vaccines [6-10]. The polymerase chain reaction (PCR) test has been introduced as one of the primary methods for detecting COVID19 [11]. However, the PCR test is a laborious, time-consuming, and complicated process with current kits in short supply [12]. On the other hand, X-ray images are extensively accessible [13-15], and scans are comparatively low-cost [16-18].

Therefore, a method based on chest X-ray imaging has become almost the most useful method to detect COVID19 positive cases [19]. However, this method suffers from the long-time needed by the radiologists to read and interpret X-ray images [20]. Besides, due to the increasing prevalence of the COVID19 virus, the number of patients, who need an X-ray image interpretation, is much higher than the number of radiologists leading to the radiologists overloaded, long-time diagnosis process, and a critical risk of other people's infection. Thereby, the rapid and automated X-ray image interpretation for accurately diagnosing the COVID19 positive cases is necessary. In this regard, Computer-Aided Diagnostic (CAD) models have been recently utilized to help radiologists [6, 8, 19, 21].

Deep Learning (DL) models have been widely utilized in various challenging image processing and classification tasks [22-24], including the COVID19 positive cases' early detection and diagnosis [25]. Deep-COVID [26] was almost the pioneer in COVID19 detection using DL models. In this research, four well-known DEEP CNNs, including SqueezeNet, ResNet18, ResNet50, and DenseNet-121 were proposed to identify COVID19 positive cases in the analyzed chest X-ray images. Aside from the results, this reference provides a unique dataset of 5000 Chest X-rays (called COVID-Xray-5k) that radiologists have validated. This distinctive feature of the provided dataset motivates us to use it as a benchmark dataset.

In [27], an automated DarkNet model was used to perform a binary and a multiclass classification task. This model has designed to achieve up to 98 % accuracy, but it used seventeen convolutional layers and numerous filtering on each layer leading to a model with high complexity. A particular deep CNN named CoroNet [28] was proposed to recognize COVID19 positive cases from chest X-ray images automatically. CoroNet is based on Xception architecture pre-trained on ImageNet dataset and trained end-to-end on a dataset developed by gathering COVID19 and other chest pneumonia X-ray images from two separate publicly accessible databases. Although the proposed model was fast and straightforward, the results were highly tolerable in accuracy and reliability. A customized deep CNN for detecting COVID19 positive cases, named COVID-Net, was proposed by [29]. This model was utilized to divide the chest X-ray image into normal and COVID19 classes. The performance of the COVID-Net model was evaluated using two publicly available datasets. It is noted that the highest accuracy rate of 92.4% was obtained by COVID-Net, which is not very interesting. COVIDX-Net [30] is another DL model utilized to diagnose the COVID19 positive cases by chest X-ray images' analysis. This model has been evaluated on seven well-known pre-trained models (e.g., DenseNet201, VGG19, ResNetV2, Inception, Xception, MobileNet, and V2InceptionV3) using a small dataset of fifty X-ray images. In this experiment, the highest accuracy rate of 91% was obtained using the DenseNet201. Reference [31] proposed a novel model to select the best COVID19 detector using the TOPSIS and Entropy technique as well as 12 machine learning classifiers. The linear SVM classifier obtained the highest accuracy of 98.99%. Although the proposed represents a high classification accuracy, the model complexity was very high in time and space.

In another point of view, several deep CNNs were also utilized as feature descriptors to transfer the input image into lower-dimensional feature vectors [32-35]. Afterward, these extracted feature vectors were fed into various classifiers to produce the final decision. Despite the reasonable classification accuracy (between 98% and 99%), these methods require manual parameter setting and matching feature extraction section with the classifier section. Also, the complexity of the final model is relatively high.

On the other hand, several methods have utilized preprocessing methods to improve the performance of classifiers. In [36], Authors tried to use preprocessing methods to eliminate diaphragms, normalize X-ray image contrast-to-noise ratio, and produce three preprocessed images, which are then linked to a transfer learning-based deep CNN (i.e., VGG16) to categorize chest X-ray images into three classes of COVID-19, pneumonia, and normal cases. The classifier obtained the highest accuracy of 93.9%. A comparison study between VGG-19, Inception_V2, and the decision tree model was performed in [8] to develop a binary classifier. In this work, first, the input images' noise level was eliminated using a feature detection kernel to produce compact feature maps. These feature maps were fed into the DL models as input. The best accuracy rate of 91% was obtained using VGG-19 compared to 78%, and 60% were obtained by Inception_V2 and the decision tree method, in order. In [37], after using a preprocessing model to detect and eliminate diaphragm areas showing on images, a histogram equalization algorithm and a bilateral filter are utilized to process the primary images to produce two sets of filtered images. Afterward, the primary image and the two filtered images are applied as inputs of three channels of the deep CNN to increase the model's learning information. The designed model with two preprocessing stages generates a total accuracy of 94.5 %, whereas without using two preprocessing steps, the designed model generates a lower classification accuracy of 88.0 %. Although these methods increase the classifier's accuracy, they will increase the overall complexity of the network.

Consequently, the necessity of designing an accurate [38-41] and real-time detector [42-44] has become more prominent. Besides, this review on COVID19 detection systems shows that most of the existing deep learning-based systems have used deep CNN-based networks [45-49]; thereby, we propose to employ the ability of deep CNN as a COVID 19 detector. However, the aforementioned CNN-based methods are time-consuming, at least throughout the training phase. Therefore, before the user obtains feedback from the training phase, training and testing time can take hours even if the detector works well in the determined case. Besides, self-learning X-ray image detection, which trains progressively based on the user's feedback, may not have an excellent user experience because it takes too long until the model converges while operating with it. In this case, the challenging point is having an appropriate model for X-ray image detection, which is efficient in both processing time and accuracy.

For the sake of having a real-time COVID19 recognizer, this paper proposes using ELM [50] instead of a fully connected layer to provide a real-time training process. In the proposed approach, we combine automatic feature learning of deep CNNs with efficient ELMs to address the mentioned shortcomings, i.e., manual feature extraction and extended training time, respectively. Consequently, the first phase is the deep CNN's training, which is considered an automatic feature extractor. In the second phase, a fully connected layer will be replaced by ELM for designing a real-time classifier.

It is proven that the ELM's origin is based on Random Vector Functional Link (RVLF) [51, 52], leading to the ultra-fast learning and outstanding generalization capability [53, 54]. Literature survey shows that ELM has been broadly utilized in many engineering applications [55-57]. Although various kinds of ELMs are now accessible for image detection and classification tasks, it confronts serious issues such as the need for many hidden nodes for better generalization and determining the activation function type. Besides, ELM's stochastic nature causes an additional uncertainty problem, particularly for high-dimensional image processing problems [58, 59].

The ELM-based models randomly select the input weights and hidden biases from which the output weights are calculated. During this procedure, ELMs try to minimize the training error and identify the smallest output weights' norm. Due to the stochastic choice of the input weights and biases in ELM, the output matrix may not indicate full column rank, leading to the system's ill-conditioned matrices that produce non-optimal solutions [60, 61]. Consequently, we apply a novel metaheuristic algorithm called SCA [62] to improve ELM conditioning and ensure optimal solutions.

For the rest of this research paper, the organization is as follows. In Section 2, some background resources are reviewed. Section 3 introduces the proposed scheme. Section 4 presents the simulation and discussion results, and finally, conclusions are presented in Section 5.

## 2. **Background and Materials.**

This section will represent the background knowledge, including the deep CNN, ELM, SCA, and *COVID-Xray-5k* dataset.

### 2.1 **Deep Convolution Neural Network**

Generally, deep CNN is a conventional Multi-Layer Perceptron (MLP) based on three concepts: connection weights sharing, local receive fields, and temporal/spatial sub-sampling [64]. These concepts can be arranged into two classes of layers, including subsampling layers and convolution layers. As shown in Fig. 1, the processing layers include three convolution layers C1, C3, and C5, located between layers S2 and S4, and final output layer F6. These sub-sampling and convolution layers are organized as feature maps. Neurons in the convolution layer are linked to a local receptive field in the prior layer. Consequently, neurons with identical feature maps (FMs) receive data from various input regions until the input is completely skimmed. However, the same weights are shared.

In the subsampling layer, the FMs are spatially by a factor of 2. As an illustration, in layer C3, the FM of size *10×10* is sub-sampled to conforming FM of size *5×5* in the next layer, S4. The classification process is the final layer (F6). Each FMs are the outcome of a convolution from the previous layer's maps by their corresponding kernel and a linear filter in this structure. The weights $w^k$ and adding bias $b^k$ generate the $k_{th}$ (FM) $FM_{ij}^k$ using the *tanh* function as Eq. (1).

$$FM_{ij}^k = \tanh((W^k \times x)_{ij} + b_k) \tag{1}$$

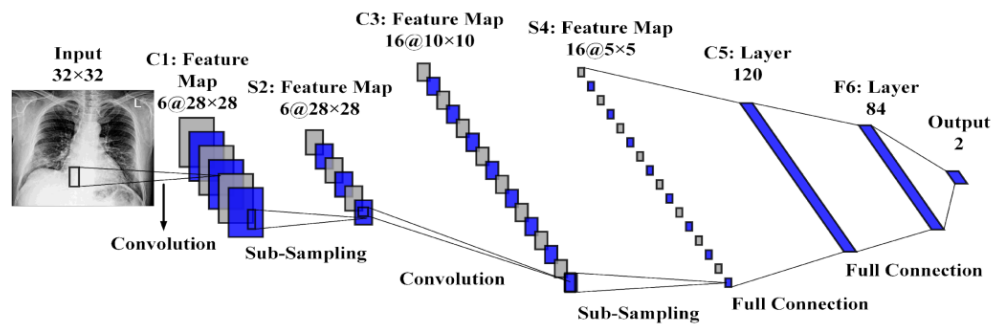

**Fig. 1** The architecture of LeNet-5 deep CNN

By reducing the resolution of FMs, the sub-sampling layer leads to spatial invariance, in which each pooled FM refers to one FM of the prior layer. The sub-sampling function is defined as Eq. (2).

$$\alpha_j = \tanh(\beta \sum_{N \times N} \alpha_i^{n \times n} + b) \tag{2}$$

Where $\alpha_i^{n \times n}$ are the inputs, $\beta$ and $b$ are trainable scalar and bias, respectively, after various convolution and sub-sampling layers. The last layer is a fully connected structure that carries out the classification task. There is one neuron for each output class. Thereby, in the case of the COVID19 dataset, this layer contains two neurons for their classes.

## 2.2. Extreme Learning Machine

ELM is one of the most widely used Single-hidden Layer Neural Network (SLNN) learning algorithms [50]. ELM first randomly sets the input layer's weights and biases and then calculates the output layer's weights using these random values. This algorithm has a faster learning rate and better performance than traditional NN algorithms. Fig. 2 indicates a typical SLNN, in which $n$ denotes the number of input layer neurons, L indicates the number of hidden layer neurons, and m shows the number of output layer neurons.

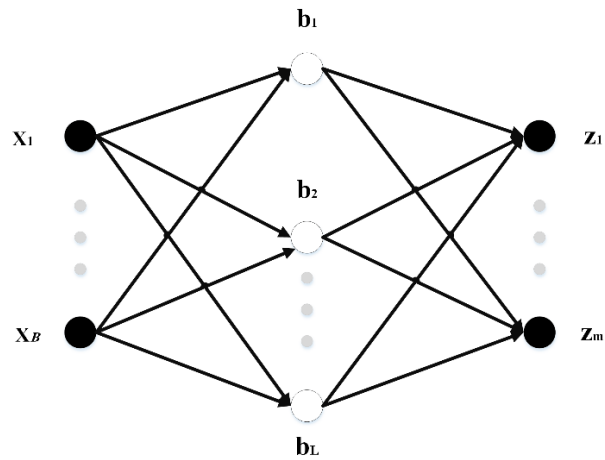

**Fig. 2** A single-hidden layer neural network.

As indicated in [50], the activation function can be shown as Eq. (3).

$$\mathbf{Z}_j = \sum_{i=1}^{L} Q_j f\left(w_i, b_i, \mathbf{x}_i\right) \tag{3}$$

Where $w_i$ denotes the input weight, $b_i$ shows the $i$th hidden neuron's bias, $\mathbf{x}_j$ represents the output weight, and $\mathbf{Z}_j$ is the SLNN output. The matrix representation of Eq. (3) is shown in Eq. (4).

$$\mathbf{Z}^T = \mathbf{H}Q \tag{4}$$

Where, $Q = [Q_1, Q_2, ..., Q_L]^T$, $\mathbf{Z}^T$ is the transpose of matrix $\mathbf{Z}$, $\mathbf{H}$ is a matrix named hidden layer output matrix, which is calculated in Eq. (5).

$$\mathbf{H} = \begin{bmatrix} f\left(w_1, b_1, \mathbf{x}_i\right) & f\left(w_2, b_2, \mathbf{x}_i\right) & \cdots & f\left(w_L, b_L, \mathbf{x}_i\right) \\ \vdots & & \cdots & \vdots \\ f\left(w_1, b_1, \mathbf{x}_\beta\right) & f\left(w_2, b_2, \mathbf{x}_\beta\right) & \cdots & f\left(w_L, b_L, \mathbf{x}_\beta\right) \end{bmatrix}_{\beta \times L} \tag{5}$$

Minimizing the training error is the primary training goal of ELM. In the conventional ELM, input biases and weights must be stochastically chosen, and the activation function must be infinitely differentiable. In this regard, the training of ELM leads to obtaining the output weight (Q) by optimizing the least-squares function indicated in Eq. (6), and the result can also be calculated as Eq. (7)

$$\min_Q \left\| \mathbf{H}Q - \mathbf{Z}^T \right\| \tag{6}$$

$$\hat{Q} = \mathbf{H}^+ \mathbf{Z}^T \tag{7}$$

In this equation, $\mathbf{H}^+$ denotes the generalized Moore-Penrose inverse of the $\mathbf{H}$ matrix.

### 2.3 Sine-Cosine Algorithm

Generally speaking, the optimization process in population-based methods begins with a series of responses that are randomly selected. The output function continually evaluates these random responses. Finally, the result of the output function gets optimized by the intended optimization method. If the number of selected responses and the iterations are appropriately considered, the probability of getting the best answer is also increased [65, 66].

Despite the differences between existing algorithms for population-based random optimization, in all of them, the optimization process is performed in two stages: exploration and exploitation [67, 68]. A randomized algorithm combines stochastic responses at a high rate in the search stage to find possible areas in search space. At the identification stage, slight changes are made to random responses, and outputs are recalculated. The method to calculate these outputs after applying changes to random responses is shown in Eqs. (8) and (9) [62].

$$X_i^{t+1} = X_i^t + r_1 \times \sin\left(r_2\right) \times \left| r_3 p_i^t - X_i^t \right| \tag{8}$$

$$X_i^{t+1} = X_i^t + r_1 \times \cos\left(r_2\right) \times \left| r_3 p_i^t - X_{ii}^t \right| \tag{9}$$

In which $X_i^t$ is the location of current response in $i$-th dimension and $t$-th iteration. Also, $r_1$, $r_2$, $r_3$ are random numbers, $p_i$ is the location of a destination in the $i$-th dimension and $|\,.\,|$ represents absolute value. Eqs. (8) and (9) can be combined to generate Eq. (10).

$$X_i^{t+1} = \begin{cases} X_i^t + r_1 \times \sin\left(r_2\right) \times \left| r_3 p_i^t - X_i^t \right| & , \ r_4 < 0.5 \\ X_i^t + r_1 \times \cos\left(r_2\right) \times \left| r_3 p_i^t - X_i^t \right| & , \ r_4 \geq 0.5 \end{cases} \tag{10}$$

In which $r_4$ is a random number in a range of [0.1]. As shown in Eq. (10), there are four main parameters $r_4$, $r_3$, $r_2$, $r_1$ in the algorithm. The parameter $r_1$ shows the next location area (or direction of motion) that can be between the source and destination (or outside of it). The parameter $r_2$ defines the amount of movement towards the destination or in the opposite direction. The parameter $r_3$ determines the size of random weight to reach the destination (which may have a value as $r_3 > 3$ or $r_3 < 3$). Eventually, $r_4$ changes equally between the components of the sinus and cosine as shown in Eq. (8). Fig. 3 shows the effect of the sinus and cosine functions on Eqs. (8) and (9). This figure shows how the proposed equation defines the area between two responses in the search area (of course, this figure is plotted for the two-dimensional space).

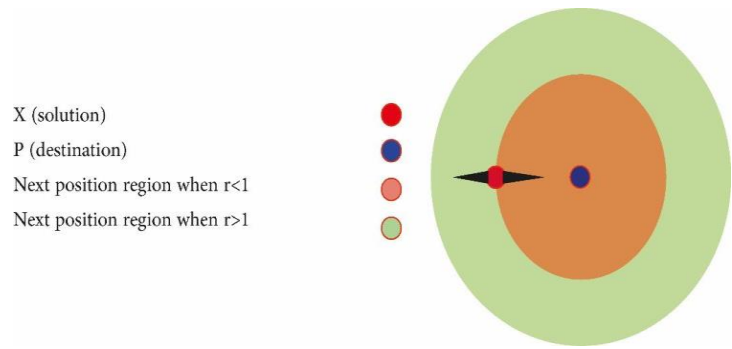

**Fig. 3** The effect of sine and cosine functions on Eqs. (8) and (9).

It should be noted, however, that Eqs. (8) and (9) can be extended to higher dimensions. The periodic form of the sinus and cosine functions allows a response to accumulating around another response. Therefore, identifying the defined space between the two responses is guaranteed. In order to find the destination (target) in the search area, the solution should search the space between similar responses (targets) comprehensively [69]. As shown in Fig. 4, this ability is achievable by changing the range of the sinus and cosine functions.

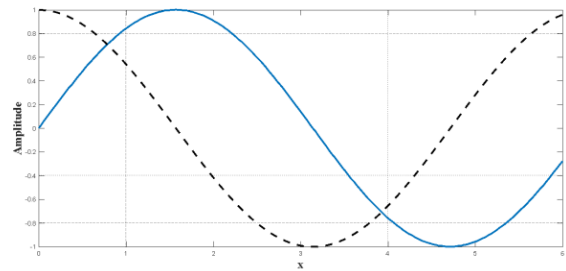

**Fig. 4** Changes in sinus and cosine functions in a specified interval

A conceptual model is shown in Fig. 5 to indicate the effectiveness of the sinus and cosine functions. This figure shows how the range of sine and cosine changes in order to update the location of a response.

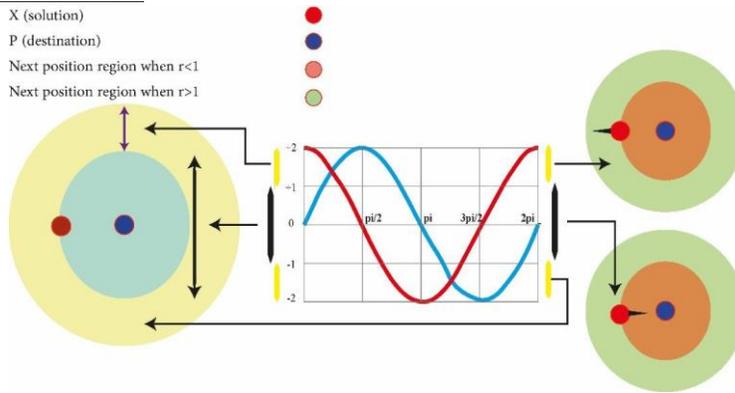

**Fig. 5** Changes in the sinus and cosine functions within the range of [-2, 2] causes to get closer or more distant from the desired response

If the parameter $r_2$ in Eq. (10) is defined as a random number in the range $[0 . 2\pi]$, then the existing mechanism guarantees to explore the search area. An appropriate algorithm should balance the exploration and exploitation operations, identify possible search areas, and ultimately converge to a general optimum. To achieve a balance between the exploitation and exploration phases, the domain of the sinus and cosine functions in Eqs. (8), (9), and (10) varies by Eq. (11).

$$r_1 = a - t \frac{a}{T} \tag{11}$$

Where $t$ is the current step, T is the maximum number of steps, and $a$ is also a fixed number. Fig. 6 shows the reduction in the range of the sinus and cosine functions during iterations.

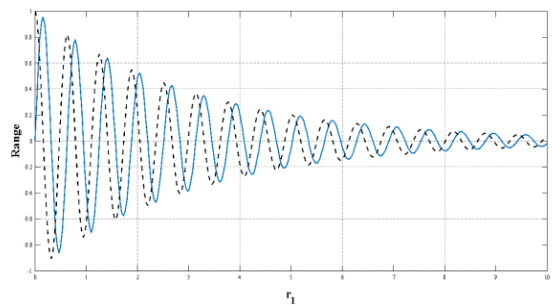

**Fig. 6** Reduction in the range of sine and cosine functions during iterations.

According to Figs. 3 and 4, when the sinus and cosine functions are in the range of [-2,-1) and (1,2], the algorithm will explore the search area. On the contrary, when they are in the range of [-1,1], the algorithm detects the search area. This figure shows that the algorithm starts the optimization process using a set of random answers. Then the algorithm reserves the best answers (solutions) that have been obtained so far. The reserved answers are set as targets, and the rest of the responses are updated according to these targets. Besides, the range of the sinus and cosine functions are updated to enhance the search space identification and increase the number of steps.

The optimization process by the algorithm ends when the number of steps exceeds the maximum defined by default. Of course, it should be noted that other conditions, such as the maximum number of function evaluations or overall optimization accuracy, can be considered as conditions to end the optimization process. By using the operators mentioned above, the proposed algorithm can solve optimization problems theoretically for the reasons given below.

- The algorithm creates and optimizes a set of random answers for a particular problem. Therefore, its advantage compared to other algorithms that are based on one response is the high exploration ability and avoidance of trapping in local minima.

- When the sinus and cosine functions have values greater than 1 or smaller than -1, different search space areas are explored to find the answer.

- When the sinus and cosine functions have values between 1 and -1, the explored areas are likely to be part of the answer.

- The algorithm alters slowly from exploration to exploitation mode based on changes in the range of the sinus and cosine functions.

- The best optimum approximation is stored in a variable as the target (response) and maintained throughout the entire optimization process.

- As responses constantly update their location around the best answer, they always tend to choose the best search area during the optimization process.

- Since the proposed algorithm considers the problem as a black box, it can be easily used for well-formulated problems.

## 2.4 **COVID-X-ray Dataset**

A dataset named *COVID-X-ray-5k* dataset, including 2084 training and 3100 test images, was utilized [26]. In this dataset, considering radiologist advice, only anterior-posterior COVID19 X-ray images are used because the lateral photos are not applicable for detection purposes. Expert radiologists evaluated those images, and those that did not have clear pieces of evidence of COVID19 were eliminated. In this way, 19 images out of the whole 203 images were removed, and 184 images remained, indicating clear pieces of evidence of COVID19. With this method, a group with a more clearly labeled dataset was introduced. Out of 184 photos, 100 images are considered for the test set, and 84 images are intended for the training set. For the sake of increasing the number of positive cases to 420, data augmentation is applied. Since the number of normal cases was small in the *covid-chestxray-dataset* [70], the supplementary *ChexPert* dataset [71] was employed. This dataset includes 224316 chest X-ray images from 65240 patients. 2000 and 3000 non-COVID images were chosen from this dataset for the training and test sets, respectively. The final number of images related to various classes is reported in Table 1. Fig. 7 indicates six stochastic sample cases from *the COVID-X-ray-5k* dataset, including two positive and four normal samples.

**Table 1** The categories of images per class in the COVID dataset.

| Category | COVID19 | Normal |
|---|---|---|
| Training Set | 420 (84 before augmentation) | 2000 |
| Test Set | 100 | 3000 |

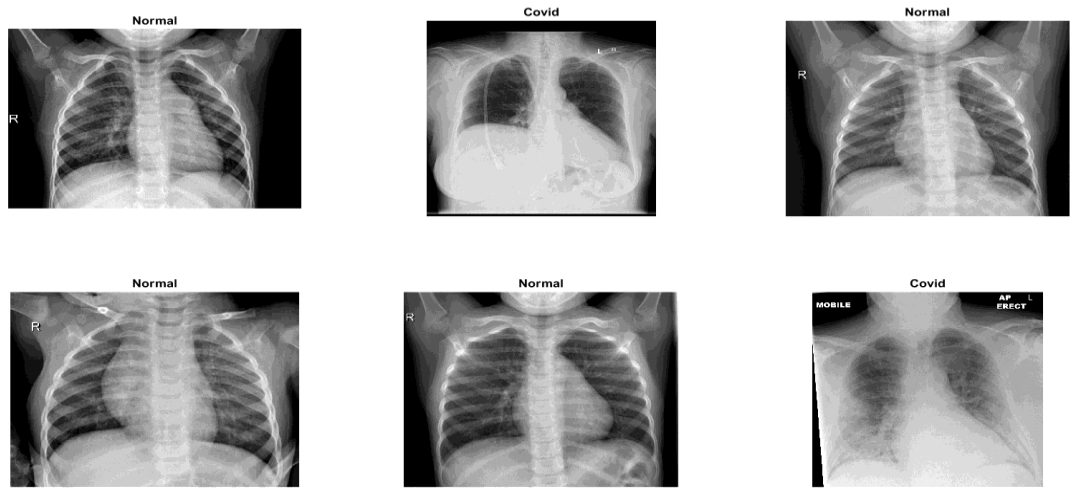

**Fig. 7** Six stochastic sample images from *the COVID-X-ray-5k* dataset

3. **Methodology**

As previously stated, this paper uses the LetNet-5 structure as a COVID19 positive cases detector. It consists of three convolutional layers, two pooling layers followed by a Fully-Connected (FC) layer, which uses Gradient Descent-based Back Propagation (GDBP) algorithm for learning. Considering the aforementioned GDBP deficiencies, we propose to use a single-layer ELM instead of FC layers to classify the extracted features, as shown in Fig. 8.

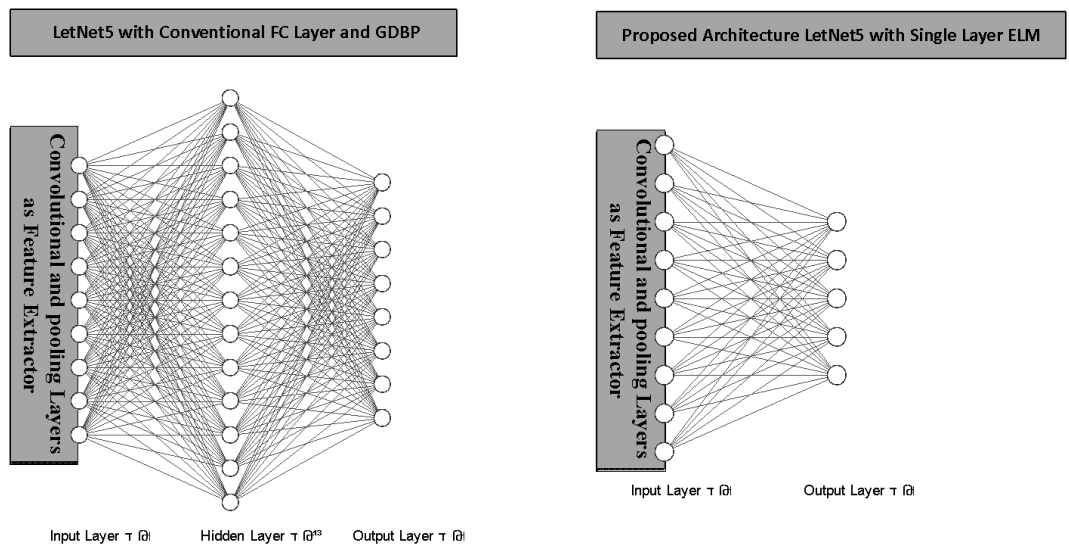

**Fig. 8** The conventional vs. proposed architecture

The convolutional layers' weights are pre-trained on a large dataset as a complete LetNet-5 with a standard GDBP learning algorithm. After the pre-training phase, the FC layers are removed, and the remaining layers are frozen to exploit as a feature extractor. The features generated by the stub-CNN will provide the ELM network's input values. In the proposed structure, ELM has 120 hidden layer neurons and two output neurons. Noted that the sigmoid function is used as an activation function.

## 3.1 Stabilizing ELM using SCA

Despite the reduction of training time in ELMs compared to the standard FC layer, ELMs are not stable and reliable in real-world engineering problems due to the random determination of the input layer's weights and biases. Thereby, we apply the SCA for tuning the input layer weights and biases of ELM to increase the network stabilization (SCA-ELM) and reliability while keeping the real-time operation. Generally, there are two main issues in tuning a deep network using a meta-heuristic optimization algorithm. First, the structure's parameters must be represented by the searching agents (candid solution) of the meta-heuristic algorithm; next, the fitness function must be defined based on the considered problem's interest.

The presentation of network parameters is a distinct phase in tuning a Deep Convolutional ELM using SCA (DCELM-SCA) algorithm. Thereby, ELM's input layer weights and biases should be determined to provide the best diagnosis accuracy. To sum up, SCA optimizes the input layer weights and biases of ELM, which are used to calculate the loss function as a fitness function. In fact, the values of weight and bias are used as searching agents in the SCA. Generally speaking, three schemes are used to present weights and biases of a DCELM as candid solutions of the meta-heuristic algorithm: vector-based, matrix-based, and binary state [72-74]. Because the SCA needs the parameters in a vector-based model, in this paper, the candid solution is shown as Eq. (12) and Fig. 9.

$$\text{Candid solution} = [W_{11}, W_{12}, ...., W_{nL}, b_1, ...., b_L] \qquad (12)$$

Where $n$ is the number of the input nodes, $W_{ij}$ indicates the connection weight between the $i_{th}$ feature node and the $j_{th}$ ELM's input neuron, and $b_j$ is the bias of the $j_{th}$ input neuron. As previously stated, the proposed architecture is a simple LeNet-5 structure [75]. In this section two structures, namely *in_6c_2p_12c_2p* and *in_8c_2p_16c_2p*, are used where *c* and *p*, are convolution and pooling layers, respectively. The kernel size of all convolution layers is 5x5, and the scale of pooling is down-sampled by a factor of 2.

## 3.2 Loss Function

In the proposed meta-heuristic method, the SCA algorithm trains DCELM to obtain the best accuracy and minimize evaluated classification error. This objective can be computed by the loss function of the metaheuristic searching agent or the Mean Square Error (MSE) of the classification procedure. However, the loss function used in this method is as follows [76]:

$$y = \frac{1}{2}\sqrt{\frac{\sum_{i=0}^{N}(\text{o}-\text{u})^2}{N}} \qquad (13)$$

Where o shows the supposed output, u indicates the desired result, and $N$ indicates the number of training samples. Two termination criteria include reaching maximum iteration or predefined loss function, are utilized by the proposed SCA algorithm. Consequently, the pseudo-code of DCELM-SCA is shown in Fig. 10. Also, a schematic workflow explaining the proposed method is shown in Fig. 11.

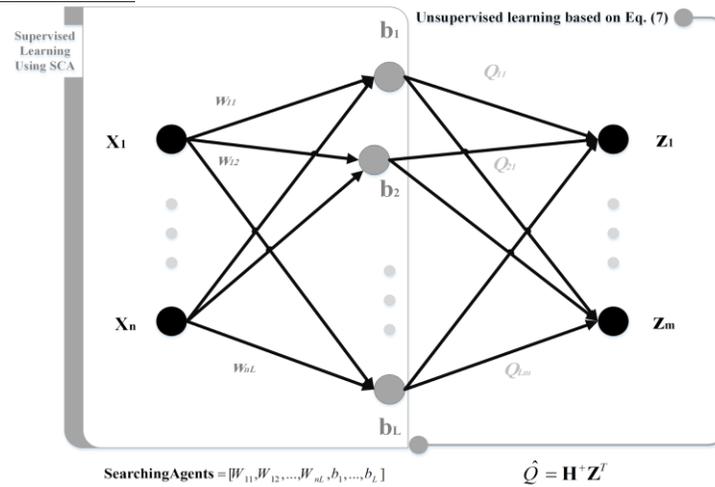

**Fig. 9** Assigning the deep CNN's parameters as the candid solution (searching agents) of SCA.

## 4. Simulation Results and Discussion

As previously stated, the hybrid method's primary target is to enhance the diagnosis rate of classic deep CNN by using the ELM and SCA learning algorithms. In the DCELM-SCA simulation, the population is equal to 50, and the maximum iteration is equal to 10. The parameter of deep CNN, i.e., the learning rate $\alpha$ and the batch size, are equal to 0.0001 and 12, respectively. Also, the number of epochs is considered between 1 and 10 for every evaluation. We down-sample all input images to 31×31 before applying them to deep CNNs. The assessment was carried out in MATLAB-R2019a on a PC with Intel Core i7-4500u processor, 16 GB RAM, in Windows 10, with ten individual runtimes. The performance of DCELM-SCA is compared with DCELM [77], DCELM-GA [78], DCELM-CS [79], and DCELM-WOA [80] on the *COVID-Xray-5k* dataset. The parameters of the SCA, GA, CS, and WOA are shown in Table 2.

| **Pseudo-Code DCELM-SCA** |
|---|
| **Result**: computation *time, and classification accuracy,* |
| *Initialization of the DCELM structure* |
| Calculation process: *loss function f(x), connection weights (w), and biases (b);* |
| **Initialize** *a set of search agents (solution) (x)* |
| **Do** |
| **Evaluate** *each of the search agents by the objective function* |
| **Update** *the best solution obtained so far (p=x*)* |
| **Update** *$r_1$, $r_2$, $r_3$ and $r_4$* |
| **Update** *the position of search agent* |
| **While** *(t < maximum number of iteration)* |
| **Return** *the best solution obtained so far as the global optimum* |

**Fig. 10** The pseudo-code for DCELM-SCA model.

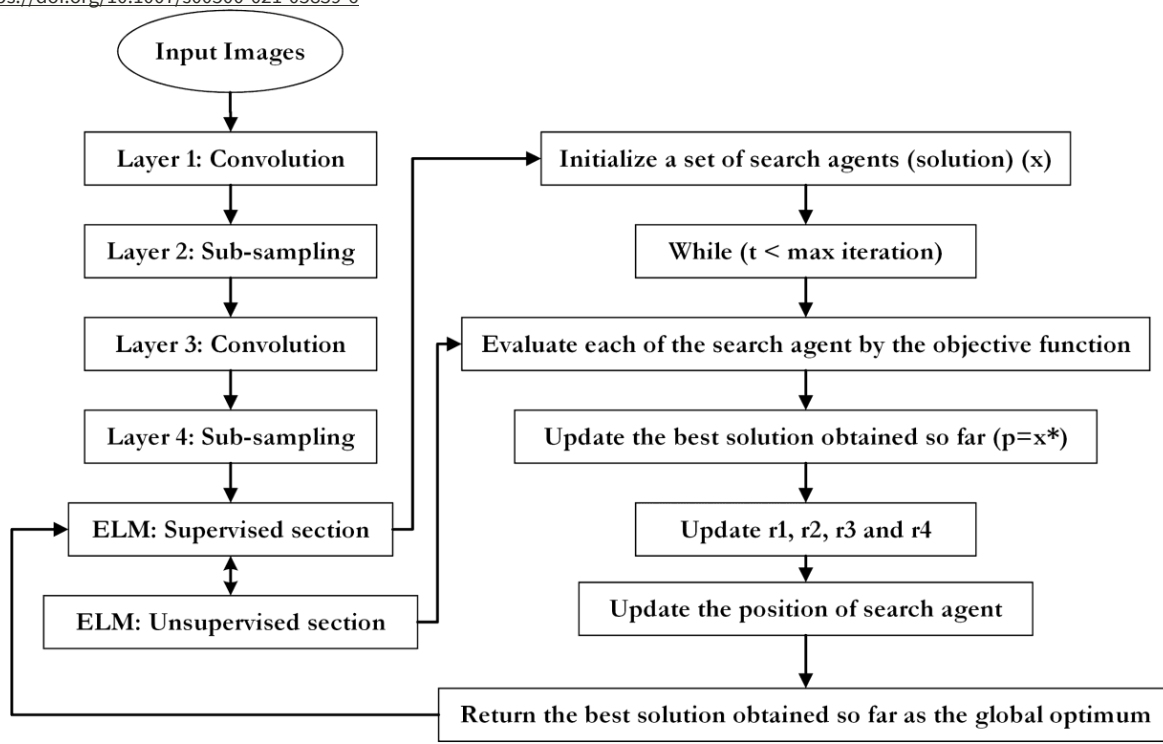

**Fig. 11** The flowchart of the designed model.

**Table 2** The parameters of benchmark algorithms

| Algorithm | Parameters | Values |
|---|---|---|
| GA | Cross-over probability | 0.7 |
| | mutation probability | 0.1 |
| | population size | 50 |
| CS | Discovery rate of alien eggs | 0.25 |
| | population size | 50 |
| WOA | $a$ | linearly decreased from 2 to 0 |
| | population size | 50 |
| SCA | $a$ | 2 |
| | $r_1$, $r_2$, $r_3$, $r_4$ | Random in the range of [0,1] |
| | population size | 50 |

### 4.1. Evaluation Metrics

Various metrics can be remarkably used to measure classification models' efficiency, such as sensitivity, classification accuracy, specificity, precision, Gmean, Norm, and F1-score. Since the dataset is significantly imbalanced (100 COVID19 images, 3000 NonCOVID images), we utilize specificity (true negative rate) and sensitivity to correctly reporting the performance of designed models, as following equations (true positive rate).

$$Sensitivity\ (TPR) = \frac{TP}{P} = \frac{TP}{TP + FN} \tag{14}$$

$$Specificity\ (TNR) = \frac{TN}{N} = \frac{TN}{TN + FP} \tag{15}$$

Where TP denotes the number of true positive cases, FN is the number of false-negative cases, TN indicates the number of true negative cases, and FP represents the number of false-positive cases.

### 4.2. Structure Expected Probability Grades

As previously stated, as the importance of time complexity, we utilize two simple LetNet-5 convolutional structures, i. e., in_6c_2p_12c_2p and in_8c_2p_16c_2p. The probability grade for each image is predicted by these structures, which indicates the possibility of the image being identified as COVID19. Comparing this likelihood with a threshold, we can extract a binary label indicating whether the specified image is COVID19 or not. A perfect structure must identify all COVID19 cases' likelihood close to one and NonCovid cases close to zero.

Fig. 12 and Fig. 13 indicate the distribution of Expected Probability Grades (EPG) for the images in the test dataset, by in_6c_2p_12c_2p and in_8c_2p_16c_2p models, respectively. Because the NonCovid category in *covid-chestxray-dataset* includes general cases and other kinds of infections, the distribution of EPG is presented for three categories, i.e., Covid19, NonCovid other infections, and NonCovid general cases. As shown in Fig. 12 and Fig. 13, the NonCovid images with other kinds of infections have slightly larger grades than the NonCovid general samples. It is logical since NonCovid other infection images are more complicated to recognize from COVID19 than general cases. Positive COVID19 cases are expected to have much higher probabilities than the NonCovid cases, which is certainly stimulating, as it indicates the structure is learning to recognize COVID19 from NonCovid samples. The confusion matrices for these two structures on *COVID-Xray-5k* are shown in Figs. 14 and 15.

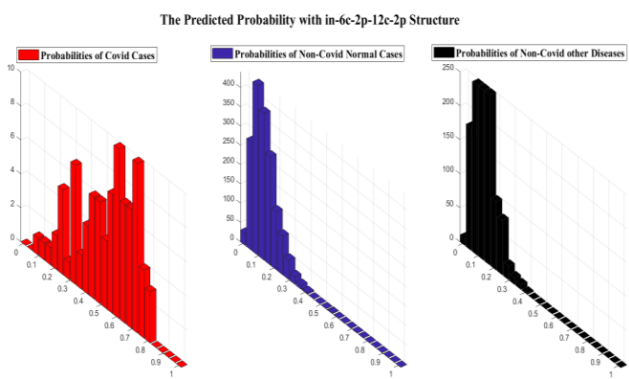

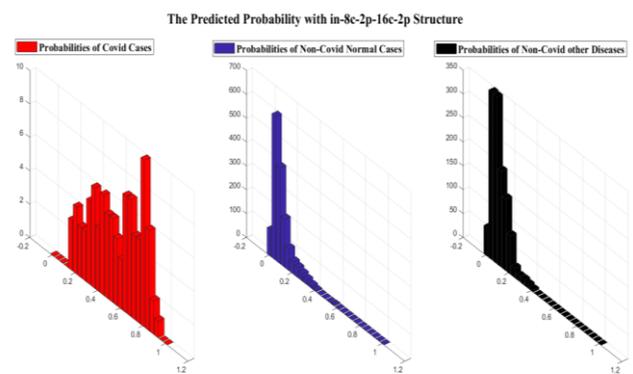

**Fig. 12** the EPG produced by in_6c_2p_12c_2p structure.     **Fig. 13** the EPG produced by in_8c_2p_16c_2p structure.

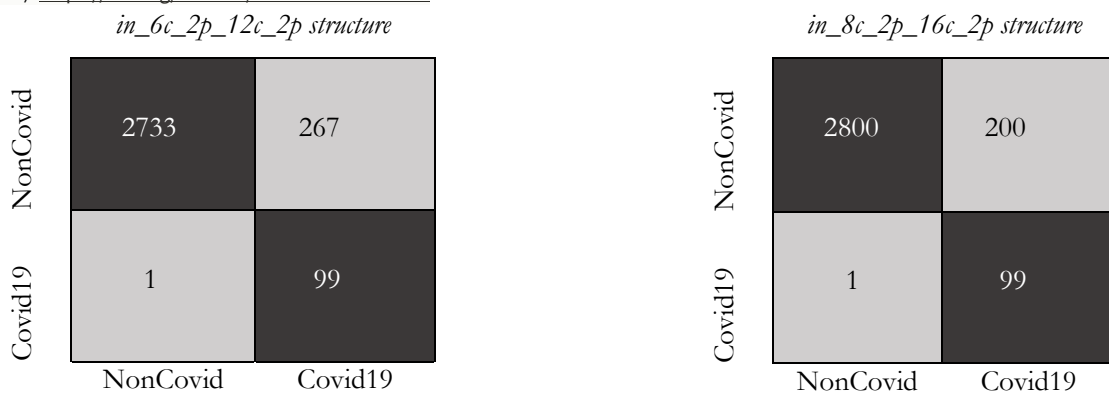

**Fig. 14** The confusion matrix for in_6c_2p_12c_2p model. **Fig. 15** The confusion matrix for in_8c_2p_16c_2p model.

Considering the calculated result, we choose the in_8c_2p_16c_2p structure as a benchmark structure named conventional deep CNN.

### 4.3. The Comparison of Specificity and Sensitivity

Each structure EPG is indicating the possibility of the image being COVID19. These EPGs can be compared with a cut-off threshold to deduce whether the image is a positive COVID19 case or not. We use calculated labels to evaluate the specificity and sensitivity of each detector. Various specificity and sensitivity rates can be calculated based on the value of the cut-off threshold. The specificity and sensitivity rates based on conventional deep CNN, DCELM, DCELM-GA, DCELM-CS, DCELM-WOA, and DCELM-SCA models for various thresholds are represented in Table 3.

Given that the results are provided for ten individual runs, Table 3 shows the Average (Ave) and Standard deviation (Std) of the results. Besides, Wilcoxon's rank-sum test [81], a non-parametric statistical test, was carried out to investigate whether the results of the DCELM-SCA differ from other compared models in a statistically significant way. It must be noted that a significance level of 5% was achieved in this case. In addition to AVE and STD, the rank-sum's p-values are reported in Table 3. It is worth noting that the N/A in the tables of results shortened form of "Not Applicable," which indicates that the relating algorithm cannot be compared with itself in Wilcoxon's test.

**Table 3** Specificity and sensitivity rates of benchmark models for various threshold values.

| Model | Threshold | Sensitivity (%) | Specificity (%) | P-value |
|---|---|---|---|---|
| | 0.1 | $98.22 \pm 0.002$ | $84.47 \pm 0.003$ | 0.0047 |
| **deep CNN** | 0.2 | $95.35 \pm 0.002$ | $85.73 \pm 0.002$ | 0.0025 |
| | 0.3 | $90.56 \pm 0.005$ | $87.42 \pm 0.002$ | 0.0047 |
| | 0.4 | $84.54 \pm 0.006$ | $90.82 \pm 0.002$ | 0.0085 |
| | 0.1 | $98.11 \pm 0.052$ | $83.37 \pm 0.082$ | 0.041 |
| | 0.2 | $94.56 \pm 0.056$ | $86.21 \pm 0.022$ | 0.0056 |
| **DCELM** | 0.3 | $89.96 \pm 0.085$ | $88.12 \pm 0.013$ | 0.0056 |
| | 0.4 | $83.22 \pm 0.101$ | $89.52 \pm 0.011$ | 3.12E-06 |
| | 0.1 | $98.33 \pm 0.002$ | $92.26 \pm 0.002$ | 0.0005 |
| | 0.2 | 97.21 ± 0.003 | 93.85 ± 0.002 | 0.002 |
| | 0.3 | 92.36 ± 0.005 | 94.85 ± 0.001 | 1.11E-11 |
| **DCELM-GA** | 0.4 | 89.24 ± 0.005 | 96.85 ± 0.001 | 0.0004 |
| | 0.1 | 99.23 ± 0.001 | 89.91 ± 0.002 | 0.0012 |
| | 0.2 | 97.63 ± 0.001 | 92.85 ± 0.002 | 0.0032 |
| **DCELM-CS** | 0.3 | 95.32 ± 0.002 | 96.33 ± 0.001 | 2.79E-06 |
| | 0.4 | 91.11 ± 0.002 | 97.33 ± 0.001 | 0.003 |
| | 0.1 | 99.01 ± 0.002 | 85.12 ± 0.004 | <u>**0.25**</u> |
| | 0.2 | 96.65 ± 0.003 | 92.98 ± 0.004 | 0.041 |
| **DCELM-WOA** | 0.3 | 91.21 ± 0.003 | 96.60 ± 0.003 | 0.045 |
| | 0.4 | 80.32 ± 0.004 | 97.90 ± 0.002 | 0.025 |
| | 0.1 | 100 ± 0.000 | 84.34 ± 0.002 | N/A |
| | 0.2 | 98.12 ± 0.001 | 93.32 ± 0.001 | N/A |
| **DCELM-SCA** | 0.3 | 97.56 ± 0.001 | 95.33 ± 0.001 | N/A |
| | 0.4 | 92.99 ± 0.002 | 98.66 ± 0.001 | N/A |

The data presented in Table 3 shows that all benchmark networks obtain very encouraging outcomes, and the best performing structure (DCELM-SCA) achieves a sensitivity rate of 100% and a specificity rate of 99.11%. In second and third place, DCELM-CS and DCELM-WOA get slightly better efficiency than other benchmark structures.

### 4.4. The Reliability Analysis of Imbalance Dataset

Considering the limitation of the number of approved labeled positive COVID19 cases, we just have 100 positive COVID19 cases in the COVID-Xray-5k dataset. Therefore, the reported sensitivity and specificity rates in Table 3 may not be completely reliable. Theoretically, more numbers of positive COVID19 cases are needed to carry out a more reliable assessment of sensitivity rates. However, the 95% confidence interval of the obtained specificity and sensitivity rates can be evaluated to examine what is the feasible interval of calculated values for the current number of test cases in each category. We can calculate the accuracy rate's confidence interval as Eq. (16) [82].

$$r = p \sqrt{\frac{\text{Accuracy.Rate}(1\text{-Accuracy.Rate})}{N}} \qquad (16)$$

Where, $p$ indicates the confidence interval's significance level, i.e., the Gaussian distribution's standard deviation, $N$ represents the number of cases for each class, Accuracy. Rate is the evaluated accuracy, which is sensitivity and specificity in this example. The 95% confidence interval is utilized to lead to the corresponding value of 1.96 for $p$. Because a sensitive network is essential for the COVID19 detection problem, the particular threshold levels are selected, which corresponds to a sensitivity rate of 98 percent for each benchmark network, and their specificity rates are then examined. The comparison of the six model's performance is presented in Table 4. The data presented in Table 4 shows that the specificity rates' confidence interval is about 1 percent. In comparison, it is equal to around 2.8 percent for sensitivity because there are 3000 images for the NonCovid class, whereas 100 images for the sensitivity rate in the test set.

**Table 4** The reliability analysis of sensitivity and specificity of four evolutionary benchmark DCELM and deep CNN.

| Model | Sensitivity (%) | Specificity (%) |
|---|---|---|
| deep CNN | 98 ± 2.8 | 84.47 ± 1.31 |
| DCELM | 98 ± 2.8 | 83.37 ± 1.32 |
| DCELM-GA | 98 ± 2.8 | 92.26 ± 0.90 |
| DCELM-CS | 98 ± 2.8 | 91.85 ± 0.91 |
| DCELM-WOA | 98 ± 2.8 | 91.33 ± 0.91 |
| DCELM-SCA | 98 ± 2.8 | 93.22 ± 0.82 |

As can be seen in Table 4, the specificity of canonical deep CNN was reduced when the ELM network was applied, i.e., the specificity of DCELM is lower than deep CNN. However, the specificity of DCELM-SCA is higher than canonical deep CNN and DCELM because of applying the SCA algorithm to improve the whole network's stability.

The comparison of various structures just based on their specificity and sensitivity rates does not represent enough information about the detector's performance because various threshold levels cause different specificity and sensitivity rates. The precision-recall curve is a good presentation that can be utilized to see the comprehensive comparison between these networks for all feasible cut-off threshold levels. This presentation indicates the precision rate as a function of recall rate. Precision is defined as the TPR divided by the TP (i.e., Eq. 14), and the recall is the same as TNR (i.e., Eq. 15). Fig. 16 shows the precision-recall plot of these six benchmark models. The Receiver Operating Characteristic (ROC) plot is another appropriate tool representing the TPR as a function of FPR. Therefore, Fig. 16 also shows the ROC curve of these six benchmark structures. The ROC curves show that DCELM-SCA significantly outperforms other DCELM based networks as well as conventional deep CNN on the test dataset. However, it should be noted that the Area Under Curve (AUC) of ROC curves may not right indicate the model efficiency since it can be very high for broadly imbalanced test sets like the COVID-Xray-5k dataset.

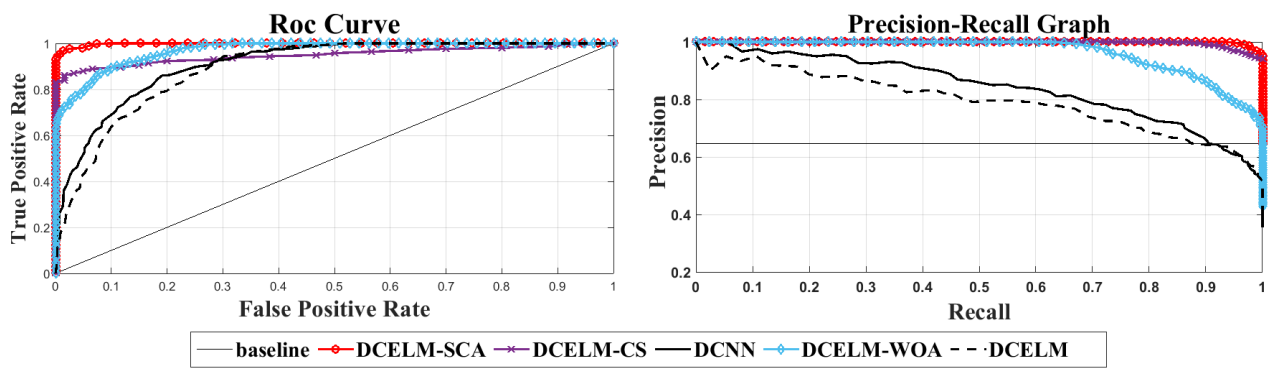

**Fig. 16** The ROC curves and Precision-recall curves for DCELM-SCA and other benchmarks.

## 4.5 The Analysis of Time Complexity

Measuring the time complexity is necessary for the sake of analyzing a real-time detector. In this regard, besides the benchmark networks, we implement the designed COVID19 detector using NVidia Tesla K20 as the GPU and an Intel Core i7-4500u processor as the CPU. The testing time is the time required to process the whole test set of 3100 images. As shown from Fig. 16, the DCELM-SCA detector indicates outstanding COVID19 detection results compared with other benchmark models. For the sake of comparison, the proposed DCELM-SCA provides over 99.11% correct COVID19 sample detection for less than a 0.89% false alarm detection rate, which shows the SCA algorithm's capability to increase the performance of the deep CNN model.

Generally, the precision-recall plot shows the tradeoff between recall and precision for various threshold levels. A high area under the precision-recall curve represents both high precision and recall, where high precision indicates a low false-positive rate, and high-recall indicates a low false-negative rate. As can be observed from the curves in Fig. 16, DCELM-SCA has a higher area under the precision-recall curves. Therefore, it means a lower false positive and false negative rate than other benchmark detectors. The simulation results indicate that DCELM-SCA represents the best accuracy for all epochs.

**Table 5** the comparison of test and training time of benchmark network implemented on GPU and CPU.

| Model | CPU vs. GPU | Training time | Testing time | P-value |
|-------|-------------|---------------|--------------|---------|
| deep CNN | GPU | 10 min, 34 sec | 3180 ms | 1.53E-07 |
| | CPU | 6 h, 44 min, 8 sec | 4 min, 30 sec | 1.37E-03 |
| DCELM | GPU | **1176 ms** | **2936 ms** | **N/A** |
| | CPU | **1 min, 26 sec** | **4 min, 19 sec** | **N/A** |
| DCELM-GA | GPU | 3645.6 ms | 3102 ms | 1.13E-03 |
| | CPU | 4 min, 26.6 sec | 4 min, 22 sec | 1.05E-04 |
| DCELM-CS | GPU | 2578.2 ms | 3101 ms | 1.62E-05 |
| | CPU | 3 min, 9.2 sec | 4 min, 27 sec | 1.32E-03 |
| DCELM-WOA | GPU | 1299.2 ms | 3015 ms | <u>0.57</u> |
| | CPU | 2 min, 9 sec | 4 min, 21sec | 1.45E-09 |
| DCELM-SCA | GPU | 1287 ms | 2985 ms | <u>**0.604**</u> |
| | CPU | 2 min, 01 sec | 4 min, 20sec | <u>**0.611**</u> |

As shown from the ROC and precision-recall curves, the Area under Curve (AUC) of DCELM (deep CNN with ELM) is reduced compared to conventional deep CNN. It means that deep CNN's performance decreases when we replace the fully-connected layer with ELM because the advantages of supervised learning are neglected. However, it is pronounced that other evolutionary deep CNNs have better performance compared to standard deep CNN. We benefit from the stochastic supervised nature of the evolutionary learning algorithm and the unsupervised nature of ELM. Consequently, the result detector's performance is improved by combining the advantages of these hybrid supervised-unsupervised learning algorithms.

From another point of view, when considering the result of Table 5, it is apparent the training and testing time of DCELMs is remarkably lower than classic deep CNN. Notably, in GPU accelerated training, the proposed approach is more than 538 times faster than the current deep CNN. Considering the number of testing and training images in Table 1 and also the entire test and training processing time in Table 5, we can easily conclude that DCELMs require less than one millisecond per image for both training and testing, thus making DCELMs real-time in both phases. Because more than 90% of the processing time is related to the feature extraction part, using other deep learning models can reduce the processing time even further.

### 4.6. Sensitivity Analysis of Designed Model

This subsection evaluates the sensitivity analysis of three control parameters employed in the designed model. The first parameter is $a$, which controls the reduction rate in the range of the sinus and cosine functions during the execution of iterations and its contribution to the optimization process, and the second and third ones are related to the network structure, i.e., the number of layers and batches. The analysis indicates which parameters are sensitive to various inputs and which ones are robust. Considering the references [84, 85], experiments were conducted by defining four-parameter levels, as represented in Table 6. Afterward, an orthogonal array can be generated to characterize various parameter combinations (as represented in Table 7). The designed model is trained for each parameter combination. The calculated MSEs for various experiments are also represented in Table 7. Considering the results from Table 7, the level trends of parameters are indicated in Fig. 17. As shown in this figure, the best performance is obtained if these three parameters are set as $N_L = 5$, $a = 1$, and $N_b = 10$.

**Table 6** the specification of parameters

| Parameters | Level | | | |
|---|---|---|---|---|
| | 1 | 2 | 3 | 4 |
| $N_l$ | 3 | 4 | 5 | 6 |
| $\alpha$ | 2.5 | 0.5 | 0.75 | 1 |
| $N_b$ | 6 | 8 | 10 | 12 |

**Table 7** Results of various parameter combinations

| Test number | Parameters | | | Result (MSE) |
|---|---|---|---|---|
| | $N_l$ | $\alpha$ | $N_b$ | |
| Ex. #1 | 3 | 0.25 | 6 | 0.1984 |
| Ex. #2 | 3 | 0.5 | 8 | 0.1652 |
| Ex. #3 | 3 | 0.75 | 10 | 0.1655 |
| Ex. #4 | 3 | 1 | 12 | 0.0952 |
| Ex. #5 | 4 | 0.25 | 8 | 0.1821 |
| Ex. #6 | 4 | 0.5 | 6 | 0.1852 |
| Ex. #7 | 4 | 0.75 | 12 | 0.1123 |
| Ex. #8 | 4 | 1 | 10 | 0.0852 |
| Ex. #9 | 5 | 0.25 | 6 | 0.0923 |
| Ex. #10 | 5 | 0.5 | 12 | 0.0601 |
| Ex. #11 | 5 | 0.75 | 8 | 0.0532 |
| Ex. #12 | 5 | 1 | 10 | 0.0423 |
| Ex. #13 | 6 | 0.25 | 12 | 0.0977 |
| Ex. #14 | 6 | 0.5 | 10 | 0.0887 |
| Ex. #15 | 6 | 0.75 | 8 | 0.0731 |
| Ex. #16 | 6 | 1 | 6 | 0.0511 |

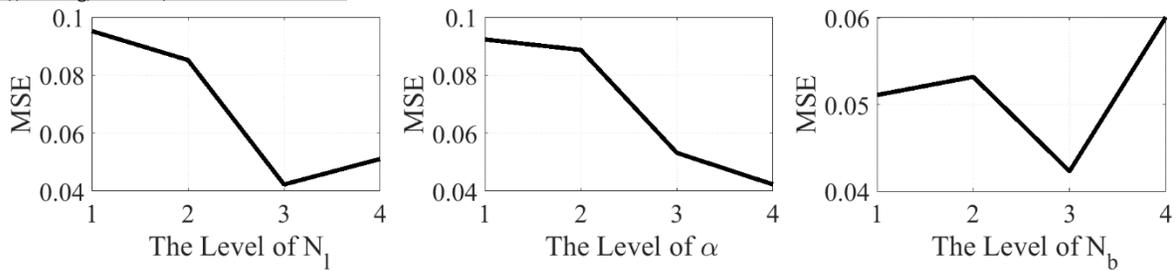

**Fig. 17** Level trends of the analyzed parameters

### 4.7. The Analysis of Convergence Behavior

For more clarification, this sub-section describes the experimental analyses of SCA's searching agents' convergence behavior. So, SCA's searching agents' convergence behavior is evaluated using qualitative metrics, including average fitness history and dynamic trajectories. Fig. 18 represents the qualitative metrics for SCA's searching agents' convergence behavior in the four categories of benchmark optimization functions (i.e., unimodal, multimodal, fixed-dimension multimodal, and composition benchmark functions), which are described in Table 8. In Fig. 18, the first column indicates the two-dimensional view of benchmark functions. The second column shows the convergence curve, which is the best solution that has been updated by now. It can be observed from the figures in this column that each group of the function represents a particular downward behavior. SCA can initially encircle the optimum point in unimodal functions and then improve the solutions as iterations pass.

Contrary, the SCA's searching agents attempt to globally discover the search space even in the final iterations to obtain superior solutions for other benchmark functions. This explorative behavior causes SCA's searching agents to the step-like convergence curves. In other words, the convergence curve indicates the performance of the best SCA's searching agents in obtaining the optimum point, whereas it does not represent any idea about the performance of the entire SCA's searching agents. For this reason, we utilize another metric to investigate the entire SCA's searching agents' performance in the optimization process named average fitness history. This metric's general pattern is almost similar to the convergence curve, while it focuses more on the total behavior and its impact on the results, improving from the initial stochastic population.

**Table 8** Benchmark functions.

| Function | Dim | Range | $f_{min}$ |
|---|---|---|---|
| **Unimodal Functions** | | | |
| $\mathrm{TF}_1(x) = \sum_{i=1}^{n} \|x_i\| + \prod_{i=1}^{n} \|x_i\|$ | 30 | $[-10,10]$ | 0 |
| $\mathrm{TF}_2(x) = \sum_{i=1}^{n-1} [100(\mathrm{x}_{i+1} - \mathrm{x}_i^2)^2 + (\mathrm{x}_i - 1)^2]$ | 30 | $[-30,30]$ | 0 |
| **Multimodal Functions** | | | |
| $\mathrm{TF}_3(\mathrm{x}) = \sum_{i=1}^{n} -x_i \sin(\sqrt{\|x_i\|})$ | 30 | $[-500,500]$ | $-418.9829 \times \mathrm{Dim}$ |
| $\mathrm{TF}_4(x) = \frac{1}{4000} \sum_{i=1}^{n} x_i^2 - \prod_{i=1}^{n} \cos\left(\frac{x_i}{\sqrt{i}}\right) + 1$ | 30 | $[-600,600]$ | 0 |
| $\mathrm{TF}_5(\mathrm{x}) = 0.1\{\sin^2(3\pi \mathrm{x}_1) + \sum_{i=1}^{n} \frac{(\mathrm{x}_i-1)^2[1+\sin^2(3\pi \mathrm{x}_i+1)]}{+(\mathrm{x}_n-1)^2[1+\sin^2(2\pi \mathrm{x}_n)]}\} + \sum_{i=1}^{n} u(x_i, 5, 100, 4)$ | 30 | $[-50,50]$ | 0 |

## Fixed-Dimension Multimodal Functions

$$\mathrm{TF}_6(x) = \left( \frac{1}{500} + \sum_{j=1}^{25} \frac{1}{j + \sum_{i=1}^{2}(x_i - a_{ij})^6} \right)^{-1} \qquad\qquad 2 \qquad [-65,65] \qquad 1$$

$$\mathrm{TF}_7(x) = [1 + (x_1 + x_2 + 1)^2 (19 - 14x_1 + 3x_1^2 - 14x_2 + 6x_1 x_2 + 3x_2^2)] \qquad 2 \qquad [-2,2] \qquad 3$$
$$\times [30 + (2x_1 - 3x_2)^2 \times (18 - 32x_1 + 12x_1^2 + 48x_2 - 36x_1 x_2 + 27x_2^2)]$$

## Composition Function

$\mathrm{TF}_8\ (CF1)$                                              10       **[−5,5]**       0
$f\,1, f\,2, f\,3, \ldots, f\,10 : Sphere\ Function$
$[\,\sigma\,1, \sigma\,2, \sigma\,3, \ldots, \sigma\,10\,] = [1, 1, 1, \ldots, 1]$
$[\,\lambda1, \lambda2, \lambda3, \ldots, \lambda10\,] = [5/100, 5/100, 5/100, \ldots, 5/100]$

$\mathrm{TF}_9\ (CF3)$                                              10       [−5,5]       0
$f\,1, f\,2, f\,3, \ldots, f\,10 : Griewank's\ Function$
$[\,\sigma\,1, \sigma\,2, \sigma\,3, \ldots, \sigma\,10\,] = [1, 1, 1, \ldots, 1]$
$[\,\lambda1, \lambda2, \lambda3, \ldots, \lambda10\,] = [1, 1, 1, \ldots, 1]$

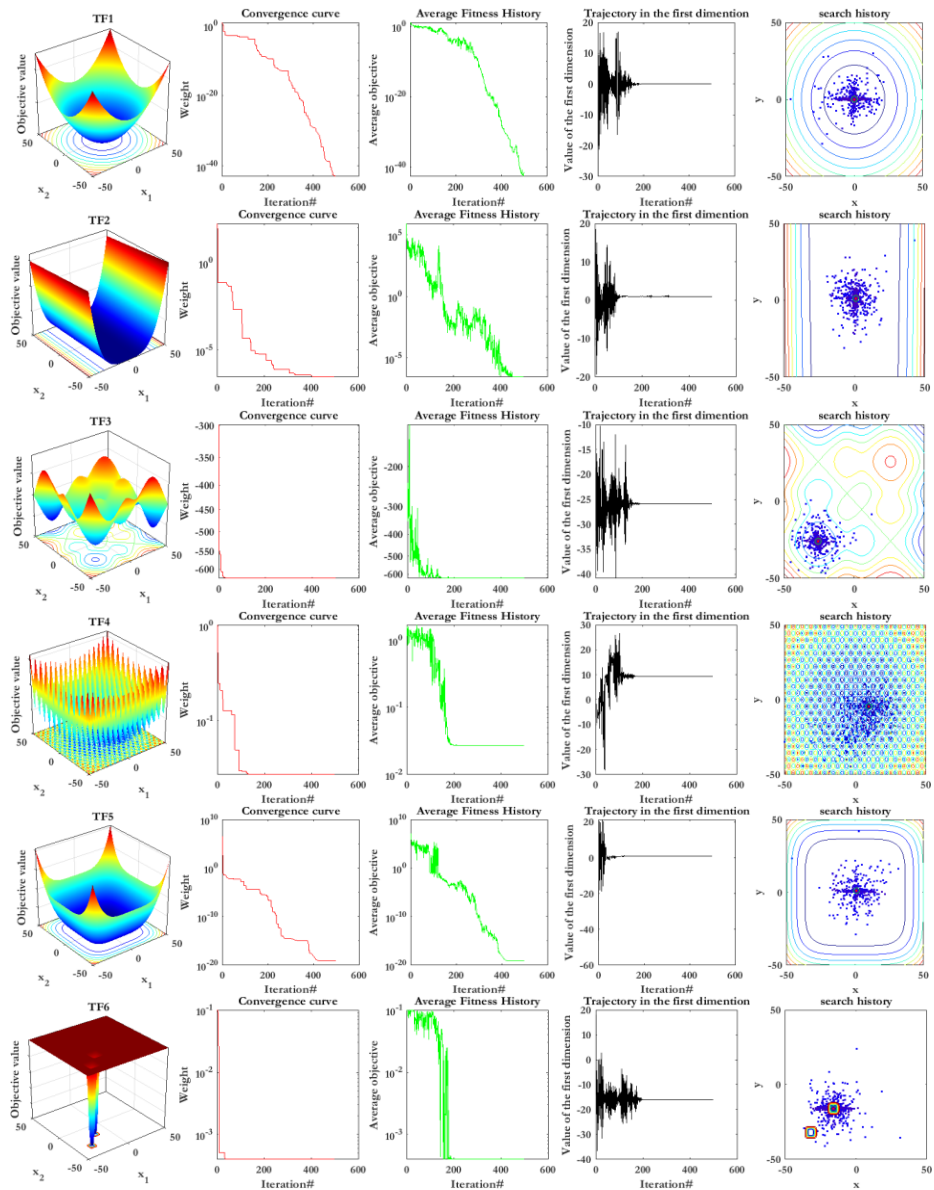

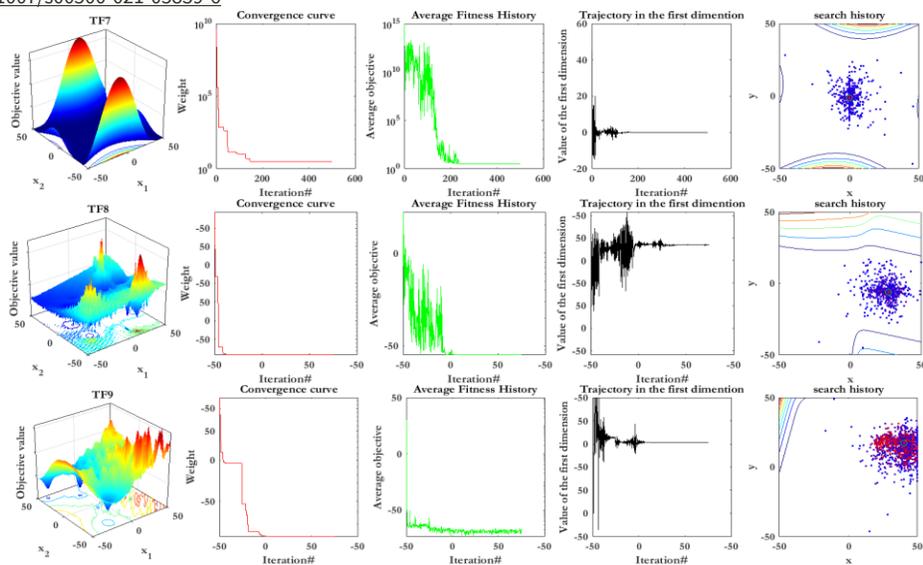

**Fig. 18** Search history, convergence curve, average fitness history, and trajectory of some functions

The trajectory of SCA's searching agents is another metric, which is represented in column four. This trajectory indicates the topological amendments from the start to the end of the optimization task. Having many dimensions in the search space, only the first dimension is selected of an agent to show its trajectory. As shown in this column, the searching agents' trajectory has high frequency and magnitude in the beginning iterations, vanishing in the last iterations. These figures verify the exploration phase in the beginning iterations while changing to the exploitation phase in the final iterations cause searching agents to converge to the global optimum finally.

The last column shows the search history as the fourth metric, indicating how searching agents' diversity causes SCA to reach global optimum among various local optima. These figures indicate a more population density around the unimodal functions' optimum points, contrary to multimodal and composition functions, in which there are more scattered SCA's searching agents in the search space.

### 4.8 Identifying the Region of Interest

From the viewpoint of data science experts, the best result could be indicated in terms of the confusion matrix, overall accuracy, precision, recall, ROC curve, etc. However, these optimal results might not be sufficient for the medical specialists and radiologists if the results cannot be interpreted. Identifying the Region of Interest (ROI) that leads to the network's decision-making will enhance the understanding of both medical specialists and data science experts.

In this section, the results provided by designed networks for the *COVID-Xray-5k* dataset were investigated. The Class Activation Mapping (CAM) [86] results were displayed for the *COVID-Xray-5k* dataset to localize the areas suspicious of the COVID19 virus. The probability predicted by the deep CNN model for each image class gets mapped back to the last convolutional layer of the corresponding model that is particular to each class to emphasize the discriminative regions. The CAM for a determined image class is the outcome of the activation map of the Rectified Linear Unit (ReLU) layer following the last convolutional layer. It is identified by how much each activation mapping contributes to the final grade of that particular class. The novelty of CAM is the total average pooling layer applied after the last convolutional layer based on the spatial location to produce the connection weights. Thereby, it permits identifying the desired regions within an X-ray image that differentiates the class specificity preceding the Softmax layer, which leads to better predictions. Demonstration using CAM for deep CNN models allows the medical specialists and radiology experts to localize the areas suspicious of the COVID19 virus, indicating Fig. 19 and Fig. 20.

Figs. 19 and 20 indicate the results for COVID19 detection in X-ray images. Fig. 19 shows the outcomes for the case marked as 'COVID19' by the radiologist, and the DCELM-SCA model predicts the same and indicates the discriminative area for its decision. Fig. 20 shows the outcomes for a 'normal' case in X-ray images, and different regions are emphasized by both comparing models for their prediction of the 'normal' subset. Now, medical specialists and radiology experts can choose the network architecture based on these decisions. This kind of CAD visualization would provide a useful second opinion to the medical specialists and radiology experts and also improve their understanding of deep learning models.

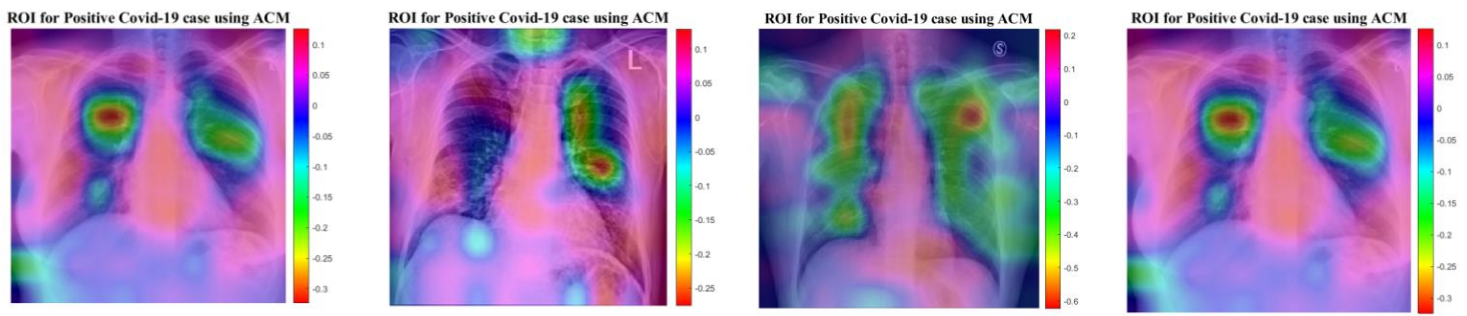

**Fig. 19** ROI for positive COVID19 cases using ACM.

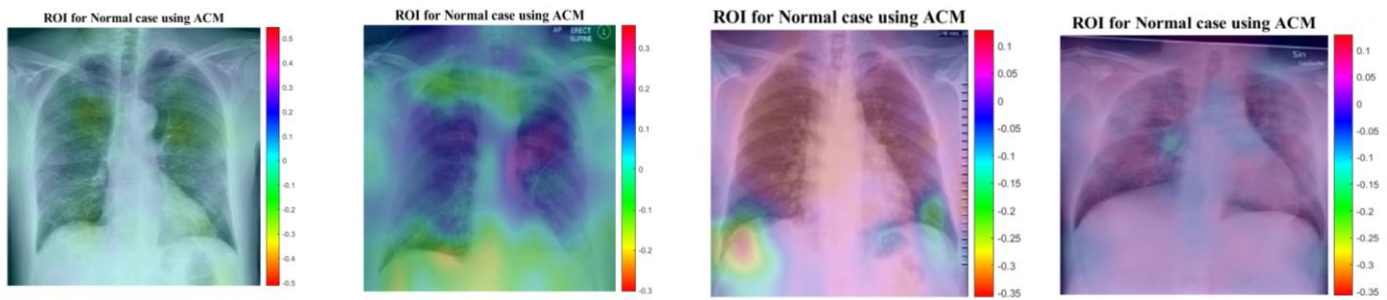

**Fig. 20** ROI for Normal cases using ACM.

## 5. **Conclusion**

In this paper, the SCA and ELM were proposed to design an accurate and reliable deep CNN model for COVID19 positive cases from X-ray images. Numerical studies were carried out to evaluate the real-time capability of the proposed model. The 95% confidence interval of the obtained specificity and sensitivity rates was performed to confirm the proposed method's reliability. According to the obtained results, we can conclude that the proposed model tends to be easier and more straightforward to implement compared to other benchmark models. Moreover, this design has the potential to be implemented in real-time COVID19 positive case detection. Consequently, we believe the proposed model and obtained numerical results are of practical interest to communities that are involved with deep neural network-based detectors and classifiers. The concept of class activation map was also applied to detect the virus's regions potentially infected. It was found to correlate with clinical results, as confirmed by experts. A few research directions can be proposed for future work with the DCELM-SCA, such as underwater sonar target detection and classification. Also, changing SCA to tackle multi-objective optimization problems can be recommended as a potential contribution. The investigation of the chaotic maps' effectiveness to improve the performance of the DCELM-SCA can be another research direction. Although the results were promising, further study is needed on a larger dataset of COVID19 images to have a more comprehensive evaluation of accuracy rates.

**Declaration**

**Funding**

Not Applicable

**Conflicts of interest/ Competing interests**

The authors declare that there is no conflict of interest

**Availability of data and material**

The resource images can be downloaded using the following link and references [87].

https://github.com/ieee8023/covid-chestxray-dataset, 2020.

**Code availability**

The source code of the models can be available by request.